\documentclass[prb,twocolumn]{revtex4}%

\newcommand{\be}{\begin{equation}}
\newcommand{\ee}{\end{equation}}
\newcommand{\bea}{\begin{eqnarray*}}
\newcommand{\eea}{\end{eqnarray*}}
\newcommand{\bean}{\begin{eqnarray}}
\newcommand{\eean}{\end{eqnarray}}

\begin{document}

\draft
\title
{\bf Thermoelectric and thermal rectification properties of quantum
dot junctions}

\author{ David M.-T. Kuo$^{1,2\dagger}$ and Yia-chung Chang$^{3*}$  }
\address{$^{1}$Department of Electrical Engineering and $^{2}$Department of Physics, National Central
University, Chungli, 320 Taiwan}

\address{$^{3}$Research Center for Applied Sciences, Academic
Sinica, Taipei, 115 Taiwan}

\date{\today}

\begin{abstract}
The electrical conductance, thermal conductance, thermal power and
figure of merit (ZT) of semiconductor quantum dots (QDs) embedded
into an insulator matrix connected with metallic electrodes are
theoretically investigated in the Coulomb blockade regime. The
multilevel Anderson model is used to simulate the multiple QDs
junction system. The charge and heat currents in the sequential
tunneling process are calculated by the Keldysh Green function
technique. In the linear response regime the ZT values are still
very impressive in the small tunneling rates case, although the
effect of electron Coulomb interaction on ZT is significant.
In the nonlinear response regime, we have demonstrated that the thermal rectification
behavior can be observed for the coupled QDs system, where the
very strong asymmetrical coupling between the dots and electrodes
, large energy level separation between dots and strong interdot
Coulomb interactions are required.
\end{abstract}

\maketitle


\section{Introduction}
Due to energy and environment issues, it becomes important to
understand the thermal properties of materials. Recently many
efforts are to seek efficient thermoelectric materials because
there exist potential applications of solid state thermal
devices.$^{1-9}$ Nevertheless, the optimization of thermoelectric
properties of materials is extremely difficult, since the figure of merit
($ZT=S^2 G_e T/\kappa$) depends on Seebeck
coefficient(S), electrical conductance ($G_e$) and thermal
conductance ($\kappa$) of the material. Tuning one of these physical quantities
will unavoidably alter the other because they are closely related.$^{7}$

Several methods were proposed to realize the enhancement of
ZT,$^{2}$ one of them is to reduce the system dimensionality.$^{8}$
Bi$_2$Te$_3$/Sb$_2$Te$_3$ superlattices$^{3}$, silicon quantum
wires$^{4}$ and PbSeTe based quantum dot (QD) superlattices$^{5}$
were experimentally demonstrated to show much higher ZT values when
compared with their corresponding bulk materials. A zero-dimension QD system was
predicted to have more pronounced enhancement in
thermoelectric efficiency due the reduced dimensionality$^{9}$.
Experimentally, it has been shown$^{5}$ that the performance of PbSeTe QDs can reach
a very impressive $ZT$ value of 2. Nevertheless, a ZT value
higher than 3 has never been reported. Note that the highest ZT
value is near 1 for conventional bulk materials.$^{1}$ Systems with
$ZT$ value larger than 3 may find application in making home refrigerators,
replacing the existing compressor-based refrigerators. In addition, they can
used in electrical power generators.$^{2}$

In order to seek a large ZT value,  a single molecular QD weakly
linked  to electrodes was proposed to exhibit an extremely large
ZT value in the Coulomb blockade regime.$^{10}$  However,
reference [10] did not take into account the molecular vibrations.
For a molecular junction, the coupling strengthes between
localized electrons and vibration modes are very strong.$^{11-17}$
Due to multiple phonon assisted processes arising from strong
electron phonon interactions, it is expected that ZT values will
be suppressed by molecular vibrations. Apart from that, such a
molecular junction is difficult to integrate with current silicon
based electronics. Therefore, we propose to use a thermoelectric device made of semiconductor QDs
embedded into amorphous insulator which has low heat conductivity. The
studied system is shown in Fig. 1. In addition, a nanoscale vacuum layer is inserted
to block the heat current delivered by phonon carriers, although it would be a challenging task to keep the vacuum layer
thin enough to allow sufficient electron tunneling. The vacuum
layer considered here can be realized using the technology similar to that used in
liquid crystal display implementation where a vacuum layer
is inserted for blocking the heat generated by the light source.

The key applications of thermoelectric devices include solid state refrigerators and
electrical generators. In solid state refrigerators (electrical
generators), one needs to remove (generate) large amount of heat
current (charge current). Consequently, a high QD density is
required for realistic applications. A single level Anderson model
can be used to simulate such a system adequately in the dilute QD-density limit.$^{10}$ However, for the
high QD-density system, one needs to consider the effect of interdot Coulomb interactions and electron hopping effect.
When QDs are embedded in an insulator matrix having a high potential barrier, electron hopping
among dots can be neglected. However, it is hard to avoid the
interdot Coulomb interactions due to its long-range tail.

In this paper, we investigate the effect of interdot Coulomb
interactions on the thermoelectric properties in the linear and
nonlinear response regimes via a multi-level Anderson
model$^{18,19}$. We found that the interdot Coulomb interactions
would suppress the ZT values and play a crucial role in determining
the thermal rectification behavior. The electrical conductivity,
thermal power, thermal conductivity and figure of merit were
typically calculated in the linear response regime, while crucial
applications of thermal devices in thermal rectifiers and
transistors require the understanding of the thermoelectric
properties in the nonlinear response regime.$^{20,21}$ The thermal
rectifiers can be used in solar energy storage and many other
applications. Therefore, it is important to take into account the
thermoelectric effects in the nonlinear regime. Here, we demonstrate
that coupled QDs can exhibit pronounced thermal rectification
behavior. Although the mechanism of thermal rectification for QD
junctions is similar to that of charge current, the heat current is
generated by temperature gradient and the consequent electrochemical
potential. It is the nonlinear relation between the applied
temperature gradient and the electrochemical potential that leads to
enhance thermal rectification behavior.

\section{Formalism}
A schematic diagram of the system of concern is shown in Fig. 1.
The Hamiltonian of the system can be described by a multi-level Anderson model:
\begin{eqnarray}
H&=&\sum_{k,\sigma,\beta} \epsilon_k
a^{\dagger}_{k,\sigma,\beta}a_{k,\sigma,\beta}+\sum_{\ell,\sigma}
E_{\ell} d^{\dagger}_{\ell,\sigma} d_{\ell,\sigma}
\nonumber \\
 &+& \sum_{\ell,\sigma}
U_{\ell} d^{\dagger}_{\ell,\sigma} d_{\ell,\sigma}
d^{\dagger}_{\ell,-\sigma} d_{\ell,-\sigma} +\frac{1}{2}\sum_{\ell
\neq j;\sigma,\sigma'}
U_{\ell,j} d^{\dagger}_{\ell,\sigma} d_{\ell,\sigma} d^{\dagger}_{j,\sigma'} d_{j,\sigma'} \\
\nonumber &+&\sum_{k,\sigma,\beta,\ell}
V_{k,\beta,\ell}a^{\dagger}_{k,\sigma,\beta}d_{\ell,\sigma}+
\sum_{k,\sigma,\beta,\ell}
V^{*}_{k,\beta,\ell}d^{\dagger}_{\ell,\sigma}a_{k,\sigma,\beta}
\end{eqnarray}
where $a^{\dagger}_{k,\sigma,\beta}$ ($a_{k,\sigma,\beta}$) creates
(destroys) an electron of momentum $k$ and spin $\sigma$ with energy
$\epsilon_k$ in the $\beta$ metallic electrode.
$d^{\dagger}_{\ell,\sigma}$ ($d_{\ell,\sigma}$) creates (destroys)
an electron with the ground-state energy $E_{\ell}$ in the $\ell$th
QD , $U_{\ell}$ and $U_{\ell,j}$ describe the intradot Coulomb
interactions and the interdot Coulomb interactions, respectively.
$V_{k,\beta,\ell}$ describes the coupling between the band states of
electrodes and the QD levels. We have ignored the excited levels of
QDs, assuming that the energy level separation between the ground
state and the first excited state within each QD is much larger than
intradot Coulomb interactions $U_{\ell}$ and thermal energy $k_BT$,
where T is the temperature of concern. We have also ignored the
inderdot hopping terms due to the high potential barrier separating
QDs. The key effects included are the intradot and interdot Coulomb
interactions and the coupling between the QDS with the metallic
leads.

Using the Keldysh-Green's function technique,$^{22,23}$ the charge
and heat currents leaving electrodes can be expressed as
\begin{eqnarray}
J_e&=&\frac{-2e}{h}\sum_{\ell} \int d\epsilon
\gamma_{\ell}(\epsilon) ImG^r_{\ell,\sigma}(\epsilon)
f_{LR}(\epsilon), \end{eqnarray}

\begin{eqnarray} &&Q\\ \nonumber
&=&\frac{-2}{h}\sum_{\ell} \int d\epsilon
\gamma_{\ell}(\epsilon)
ImG^r_{\ell,\sigma}(\epsilon)(\epsilon-E_F-e\Delta V)
f_{LR}(\epsilon) , \end{eqnarray} where the transmission factor is
$\gamma_{\ell}(\epsilon)=\frac{\Gamma_{\ell,L}(\epsilon)
\Gamma_{\ell,R}(\epsilon)}
{\Gamma_{\ell,L}(\epsilon)+\Gamma_{\ell,R}(\epsilon)}$.
$f_{LR}(\epsilon)=f_L(\epsilon)-f_R(\epsilon)$, where
$f_{L(R)}(\epsilon)=1/[e^{(\epsilon-\mu_{L(R)})/k_BT_{L(R)}}+1]$ is
the Fermi distribution functions for the left (right) electrode .
The chemical potential difference between these two electrodes is
related to the bias difference via $\mu_{L}-\mu_{R}=e \Delta V$.
$E_F$ is the Fermi energy of electrodes. $\Gamma_{\ell,L}(\epsilon)$
and $\Gamma_{\ell,R}(\epsilon)$ [$\Gamma_{\ell,\beta}=2\pi\sum_{{\bf
k}} |V_{\ell,\beta,{\bf k}}|^2 \delta(\epsilon-\epsilon_{{\bf k}})]$
denote the tunneling rates from the QDs to the left and right
electrodes, respectively. $e$ and $h$ denote the electron charge and
Plank's constant, respectively. For simplicity, these tunneling
rates will be assumed energy- and bias-independent. Therefore, the
calculation of tunneling current and heat current is entirely
determined by the spectral function,
$A(\epsilon)=ImG^r_{\ell,\sigma}(\epsilon)$, which is the imaginary
part of the retarded Green's function $G^r_{\ell,\sigma}(\epsilon)$.
The expression of retarded Green function is given by[18,19]

\begin{eqnarray}
G^{r}_{\ell,\sigma}(\epsilon)&=&(1-N_{\ell.-\sigma})\sum^{3^{n-1}}_{m=1}
\frac{p_m}{\epsilon-E_{\ell}-\Pi_m+i\Gamma_{\ell}} \nonumber \\
 &+& N_{\ell.-\sigma}\sum^{3^{n-1}}_{m=1}
\frac{p_m}{\epsilon-E_{\ell}-U_{\ell}-\Pi_m+i\Gamma_{\ell}},
\end{eqnarray}
where $n$ denotes the number of coupled QDs in each cell considered. $\Pi_m$ denotes the sum of
Coulomb interactions seen by a particle in dot $\ell$ due to other
particles in the dot $j (j\ne \ell)$, which can be occupied by
zero, one or two particles. $p_m$ denotes the probability of such
configurations.  For a three-QD cell ($\ell \neq j \neq j'$),
there are nine ($3\times 3$) configurations, and the probability
factors become $p_1=a_j a_{j'}$, $p_2=b_j a_{j'}$, $p_3=a_{j}
b_{j'}$, $p_4=c_j a_{j'}$, $p_5=c_{j'}a_{j}$, $p_6=b_j b_{j'}$,
$p_7=c_j b_{j'}$, $p_8=c_{j'} b_{j}$, and $p_9=c_{j} c_{j'}$,
where $a_j=1-(N_{j,\sigma}+N_{j,-\sigma})+c_j$,
$b_j=(N_{j,\sigma}+N_{j,-\sigma})-2c_j$, and $c_j=\langle
n_{j,-\sigma}n_{j,\sigma}\rangle$ is the intradot two-particle
correlation function. $N_{j,\sigma}$ is one particle occupation
number. Interdot Coulomb interaction factors are $\Pi_1=0$,
$\Pi_2=U_{\ell,j}$, $\Pi_3=U_{\ell,j'}$, $\Pi_4=2U_{\ell,j}$,
$\Pi_5=2U_{\ell,j'}$, $\Pi_6=U_{\ell,j}+U_{\ell,j'}$,
$\Pi_7=2U_{\ell,j}+U_{\ell,j'}$, $\Pi_8=2U_{\ell,j'}+U_{\ell,j}$,
and $\Pi_9=2U_{\ell,j}+2U_{\ell,j'}$.
$\Gamma_{\ell}=(\Gamma_{\ell,L}+\Gamma_{\ell,R})/2$ arises from
the self-energy due to the weak coupling between the QDs with
metallic leads, where the real part of self energy is ignored.
Such a self energy (ignoring the effect of electron Coulomb
interactions) is adequate within the Coulomb blockade regime,
but it does not capture the Kondo effect. The sum of probability factors $p_m$ for
all configurations is equal to 1, reflecting the fact that
$G^{r}_{\ell,\sigma}(\epsilon)$ satisfies the sum rule.

According to the expression of retarded Green's function of Eq.
(4), we need to know the single-particle and two-particle occupation numbers,
$N_{\ell,\sigma}(N_{\ell,-\sigma})$  and $N_{\ell,\ell}=c_{\ell}$,
which can be obtained by solving the following equations
self-consistently.
\begin{small}
\begin{equation}
N_{\ell,\sigma} = -\int \frac{d\epsilon}{\pi}
\frac{\Gamma_{\ell,L} f_{L}(\epsilon)+\Gamma_{\ell,R}
f_{R}(\epsilon)}{\Gamma_{\ell,L}+\Gamma_{\ell,R}}
ImG^r_{\ell,\sigma}(\epsilon),
\end{equation}
\end{small}
\begin{eqnarray}
c_{\ell} = -\int \frac{d\epsilon}{\pi} \frac{\Gamma_{\ell,L}
f_{L}(\epsilon)+\Gamma_{\ell,R}
f_{R}(\epsilon)}{\Gamma_{\ell,L}+\Gamma_{\ell,R}} Im
G^{r}_{\ell,\ell}(\epsilon).
\end{eqnarray}
The values of $N_{\ell,\sigma}$ and $c_{\ell}$ are restricted
between 0 and 1. The expression of two particle retarded Green
function of Eq. (6) is

\[
G^r_{\ell,\ell}(\epsilon)=N_{\ell.-\sigma}\sum^{3^{n-1}}_{m=1}
\frac{p_m}{\epsilon-E_{\ell}-U_{\ell}-\Pi_m+i\Gamma_{\ell}}.
\]

\section{Linear regime}
In the linear response regime, Eqs. (2) and (3) can be rewritten as
\begin{eqnarray}
J_e&=&{\cal L}_{11} \frac{\Delta V}{T}+{\cal L}_{12} \frac{\Delta
T}{T^2}\\ \nonumber Q&=&{\cal L}_{21} \frac{\Delta V}{T}+{\cal
L}_{22} \frac{\Delta T}{T^2},
\end{eqnarray}
where $\Delta T=T_L-T_R$ is the temperature difference across the junction.
Coefficients in Eq. (7) are given by
\begin{equation}
{\cal L}_{11}=\frac{2e^2T}{h} \int d\epsilon {\cal T}(\epsilon)
(\frac{\partial f(\epsilon)}{\partial E_F})_T,
\end{equation}
\begin{equation}
{\cal L}_{12}=\frac{2eT^2}{h} \int d\epsilon {\cal T}(\epsilon)
(\frac{\partial f(\epsilon)}{\partial T})_{E_F},
\end{equation}

\begin{equation}
{\cal L}_{21}=\frac{2eT}{h} \int d\epsilon {\cal
T}(\epsilon)(\epsilon-E_F) (\frac{\partial f(\epsilon)}{\partial
E_F})_T,
\end{equation}
and
\begin{equation}
{\cal L}_{22}=\frac{2T^2}{h} \int d\epsilon {\cal T}(\epsilon)
(\epsilon-E_F)(\frac{\partial f(\epsilon)}{\partial T})_{E_F}.
\end{equation}
Here ${\cal T}(\epsilon)=-\sum_{\ell}
\frac{\Gamma_{\ell,L}(\epsilon) \Gamma_{\ell,R}(\epsilon)}
{\Gamma_{\ell,L}(\epsilon)+\Gamma_{\ell,R}(\epsilon)}ImG^r_{\ell,\sigma}(\epsilon)|_{\Delta
V=0, \Delta T=0}$ and
$f(\epsilon)=1/[e^{(\epsilon-E_F)/k_BT}+1]$. Note that the
Onsager relation ${\cal L}_{12}={\cal L}_{21}$ is preserved. Based
on Eq. (7), the charge current can be generated by the voltage
difference and temperature gradient. If the system is in an open
circuit, the electrochemical potential will form in response to a
temperature gradient; this electrochemical potential is known as
the Seebeck voltage (Seebeck effect). Seebeck coefficient ( the
amount of voltage generated  per unit temperature gradient) is
defined as $S=\frac{\Delta V}{\Delta T}=-\frac{1}{T} \frac{{\cal L
}_{12}}{{\cal L}_{11}}$. In terms of the Seebeck coefficient, the
electron thermal conductance is  $\kappa_e= (\frac{{\cal
L}_{22}}{T^2}-{\cal L}_{11}S^2)$. To judge whether the system is able
to  generate power or refrigerate efficiently, we need to evaluate the figure of merit, $ZT=S^2 G_e
T/\kappa$, where $G_e=\frac{1}{T} {\cal L}_{11}$ is the electrical
conductance and $\kappa=\kappa_e+\kappa_{ph}$ is the thermal
conductance. $\kappa_{ph}$ denotes the thermal conductance due to the phonon contribution.
. For a system with an efficient thermoelectric properties we want
ZT as high as possible. This implies that we desire a system with
high Seebeck coefficient, high electrical conductance and low
thermal conductance. The thermal conductance arising from phonons
can be neglected ($\kappa=\kappa_{e}$) in our proposed system because the vacuum layer can
block the heat current carried by phonons effectively.

Although Eq. (4) can be employed to calculate the charge current
and heat current of a junction system with arbitrary QD
number,$^{18,19}$ here we use the three-QD example to investigate the
effect of interdot Coulomb interaction on the figure of merit, ZT. As mentioned above, ZT
depends on the electrical conductance $G_e$, Seebeck coefficient $S$
and electron thermal conductance $\kappa_e$. Therefore, it is
difficult to calculate the exact solution of ZT for arbitrary
parameters. For simplicity, we have ignored the QD size
fluctuations and assumed all QDs have the same ground-state energy, $E_{\ell}=E_g$ in the evaluation of ZT.
The QD size fluctuations will become important in the consideration of
heat current rectification below. TClosed form expressions for the coefficients defined in Eqs. (8)-(11)
exist within the small tunneling-rate limit (i.e.
$\frac{\Gamma/2}{(\epsilon-E_g)^2+\Gamma^2/4}$ can be approximated by $\pi
\delta(\epsilon-E_g)$) and no electron Coulomb interaction.
We obtain
\[ {\cal L}_{11}=\alpha_0
/\cosh^2(\Delta/(2k_BT)),\] \[{\cal L}_{12}={\cal L}_{21}=\alpha_1
/\cosh^2(\Delta/(2k_BT)),\] and \[{\cal L}_{22}=\alpha_2
/\cosh^2(\Delta/(2k_BT)),\] where
$\Delta\equiv E_g-E_F$, $\alpha_0=\frac{3e^2\pi}{2hk_B}\frac{\Gamma_L
\Gamma_R}{\Gamma}$, $\alpha_1=\frac{3e\pi}{2hk_B}\frac{\Gamma_L
\Gamma_R}{\Gamma}\Delta$, and
$\alpha_2=\frac{3\pi}{2hk_B}\frac{\Gamma_L
\Gamma_R}{\Gamma}\Delta^2$. We find that the thermal conductance
$\kappa_e= (\frac{{\cal L}_{22}}{T^2}-{\cal L}_{11}S^2)$ vanishes,
whereas the electrical conductance of $G_e=\frac{1}{T} {\cal
L}_{11}$ and the thermal power of $S=-\frac{1}{T} \frac{{\cal L
}_{12}}{{\cal L}_{11}}$ remain finite. This indicates that system ZT
diverges as $\Gamma$ approaches zero. This is the so called "Carnot
efficiency".$^{10}$

Closed-form expressions for these coefficients for finite $\Gamma$ in the
non-interacting case have also been derived in terms of trigamma functions.$^{10}$
However, the complicated trigamma functions do not simplify the expression of ZT and
make it difficult to elucidate mechanisms for optimizing ZT. Therefore, we
numerically calculate the figure of merit $ZT$ with and without the Coulomb interactions. We first consider the
case of symmetrical tunneling rates ($\Gamma_L=\Gamma_R=1 meV$).
The consideration of asymmetrical tunneling rates is not important for the linear response regime,
but it is crucial for the nonlinear response
regime.$^{20,21}$ which we shall address in the next section. ZT as a function of temperature for various values of
$\Delta$ in the absence of interdot Coulomb interactions is shown
in Fig. 2. Solid lines and dotted lines denote cases without and with
intradot Coulomb interaction $(U=125\Gamma)$, respectively. Note
that all energies are measured in terms of $\Gamma$
through out this article. We see that the solid lines merge
with the dashed lines at low temperatures. This indicates
that the effect of intradot Coulomb interaction on ZT can be ignored
when $U \gg k_BT$. Such a result can be understood as
follows. When interdot Coulomb interactions $U_{\ell,j}=U_{ds}$ vanish, the
retarded Green function consists of two branches

\begin{equation}
G^{r}_{\ell}(\epsilon)=\frac{1-N_{\ell,\sigma}}{\epsilon-E_g+\Gamma}
+\frac{N_{\ell,\sigma}}{\epsilon-E_g-U+\Gamma}.
\end{equation}
The second branch has a negligible contribution due to the vanishing factor
$exp^{-(E_g+U-E_F)/k_BT}$ (when $U \gg k_BT$ and $\Delta > 0$) which appears in Eqs. (8)-(11). The factor
$(1-N_{\ell,\sigma})$ in the first branch of Eq. (12) only affects the coefficients
(${\cal L}_{11}$, ${\cal L}_{12}$, ${\cal L}_{21}$ and ${\cal
L}_{22}$), but not their ratios. This explains why
$ZT(U=0)\approx ZT(U\gg k_BT)$. The reduction of ZT at
finite U (for instance $U/k_BT = 5$) can be understood as follows. In the small
tunneling rate limit, we find that $S^2G_e \propto
\Gamma$ and $\kappa_e \propto \Gamma$ for finite $U$. This is different from the behavior, $S^2 G_e \propto \Gamma$
and $\kappa_e \propto \Gamma^2$ in the absence of U. Consequently,
the reduction of ZT is observed in Fig. 2.

For a thermal electric device with high QD density, the interdot Coulomb interactions are also important.
Fig. 3 shows the ZT value as a function of temperature for a three-QD cell for various QD configurations
with $\Delta=30~\Gamma$ and $U=125~\Gamma$. Dotted line denotes the case of dilute QD density.
As a result of a large separation between QDs, the
interdot Coulomb interactions are negligible ($U_{\ell,j}=0$).
Dashed lines denotes the case where dot A and dot B are close to each other
($U_{AB}=45\Gamma$), but dot C is far from them
($U_{AC}=U_{BC}=0$). Dot-dashed line denotes the case with $U_{AB}=U_{AC}=45\Gamma$ but
$U_{BC}=0$. Solid line denotes the case with $U_{AB}=U_{AC}=U_{BC}=45\Gamma$. The results of
Fig. 3 indicate considerable reduction of ZT at high temperatures due to the interdot Coulomb interactions (proximity effect).
However, the proximity effect on ZT
can be ignored when $U_{\ell,j}/k_BT \gg 1$. In general, the
maximum values of interdot Coulomb interactions are one-half of
intradot Coulomb interactions. For silicon QDs embedded in
$SiO_2$, the Si QDs with $5-10$ nm diameters have the intradot
Coulomb interaction strengthes between $100-150~$meV. Therefore,
the condition of $U_{\ell,j}/k_BT \gg 1$ is not easy to be
satisfied at room temperature.

According to the results of Fig. 2, the system ZT can be tuned by
the $\Delta$ value.  In Fig. 4, we plot ZT as a function of $\Delta$
for the three-QD system with and without Coulomb interactions for
various temperatures. For the noninteracting case (thin solid line
with mark), the maximum ZT value occurred at $\Delta_{max}=2.4
k_BT$, which has been pointed out in Ref. 10. However, for the case
with finite electron Coulomb interactions, we found
$\Delta_{max,1}=3.1k_BT$, $\Delta_{max,2}=3k_BT$ and
$\Delta_{max,3}=2.8k_BT$. It is worth noting that the maximum ZT
values for different temperatures still reach $ZT_{max} \ge 3$,
which are very encouraging values. However, we have not considered
the QD size fluctuations, defects between metallic electrodes and
insulators, and electron-phonon interactions. In order to include
these affects fully, we phenomenologically replace the imaginary
part of retarded Green function of Eq. (4) by
$\Gamma_{\ell}+\Gamma_{ie}$. This means the total level-width is
expressed  as the sum of elastic and inelastic widths. Fig. 5 shows
the inelastic scattering effect on ZT. For $\Gamma_{ie}=3\Gamma$,
the maximum ZT value becomes smaller than 4. The results of Fig. 5
indicate that the suppression of $ZT_{max}$ resulting from the
inelastic scattering is serious.

To further understand the results of Figs. 3 and 4, we analyze the electron conductance $G_e$, thermal
power S and electron thermal conductance $\kappa_e$ of the system. Fig. 6 shows $G_e$, S
and $\kappa_e$ as functions of temperature for a three-QD cell for various
configurations: $U_{AB}=U_{AC}=U_{BC}=45\Gamma$ (dotted curves), $U_{AB}=U_{AC}=45\Gamma$
and $U_{BC}=0$ (dashed curves), and $U_{AB}=45\Gamma$ and $U_{AC}=U_{BC}=0 $ (solid curves),
which correspond to strong, medium, and weak proximity effects.
We noticed from Fig. 6(b) that the thermal power (S) is not sensitive to the proximity
effect, which means the proximity effects on ${\cal L}_{12}$ and  ${\cal
L}_{11}$ are similar.  Thus, the ZT behavior at high temperature shown in Fig. 3 is
mainly attributed to $\kappa_e$ and $G_e$. $\kappa_e$ is enhanced [see Fig. 6(c)],
but $G_e$ is suppressed [see Fig. 6(a)] at high temperature when the proximity
effect increases. This explains why ZT is suppressed at high
temperature with increasing proximity effect.  The maximum absolute value of S appears at near $k_BT=6\Gamma$,
whereas the maximum ZT value shown in Fig. 3 appears between
$k_BT=10\Gamma$ and $k_BT=15\Gamma$. So the temperature dependence of ZT
is similar to that of the electrical conductivity $G_e$, meaning that  $S^2 T/\kappa_e$ has a weak
temperature dependence.

In Fig. 4, we have tuned $\Delta$ from 0 to $120\Gamma$. This
implies that the energy levels of QDs are shifted away from the
Fermi energy of electrodes. ZT becomes small when $\Delta/k_BT \gg
1$. We can apply a gate voltage ($V_g$) to move $E_g$ relative to $E_F$. In. Fig. 7 we
plot $G_e$, S and $\kappa_e$ as functions of the gate voltage ($V_g$) for
various temperatures with $\Delta$ fixed at $30\Gamma$. The electric
conductance ($G_e$) clearly exhibits a Coulomb oscillation arising from
the intradot and interdot Coulomb interactions. The first three
peaks of $G_e$ result from the resonant channels of poles at $E_g$,
$E_g+U_{ds}$, and $E_g+2U_{ds}$ for $\epsilon$. Other peaks can be readily identified by the resonant
channels of retarded Green function of Eq. (4). Note that the
resonant channel of $\epsilon=E_g+U$ (= $155 \Gamma$) is seriously suppressed due
to the fact that all three QDs are filled with one electron at that gate voltage. The
Coulomb oscillatory behavior of $G_e$ becomes smeared at higher temperatures.

The thermal power ($S$) exhibits a sawtooth-like shape with respect
to gate voltage, which is consistent with the experimental
observation.$^{24}$ The sawtooth-like shape was also theoretically
reported in the metallic single electron transistor,$^{25,26}$ where
the charging energies are homogeneous. In Refs. 25 and 26, a model
based on the rate equations was adopted. The thermal power $S$ can
be tuned from negative to positive values. When the Fermi energy
matches a resonant channel ($G_e$ reaches a maximum value) the
thermal power vanishes, since ${\cal L}_{12}=0$. When the Fermi
energy of electrodes is in the middle of two resonant channels, $S$
also vanishes. Zero thermal power indicates that the current arising
from temperature gradient can be self-consistently balanced without
electrochemical potential. The behavior of $\kappa_e$ is much more
complicated than that of $G_e$, since $\kappa_e= (\frac{{\cal
L}_{22}}{T^2}-{\cal L}_{11}S^2)$ which consists of ${\cal L}_{11}$,
${\cal L}_{12}$, and ${\cal L}_{22}$. Since $\kappa_e$ is positive
definite, we obtain the relation $\frac{{\cal L}_{22}}{T^2} \ge
{\cal L}_{11}S^2$. When thermal power vanishes, $\kappa_e =
\frac{{\cal L}_{22}}{T^2}$. Based on the results shown in Fig. 7,
the optimized ZT value does not match either the maximum $G_e$ (good
conductor) or the minimum $G_e$ (poor conductor). The largest value
for ZT is obtained midway between the good and poor conductors as
illustrated in Fig. 8 for $k_BT=2\Gamma$ and $\Delta=30\Gamma$. So
far, our discussion is limited to the linear response regime with
$\Delta T/T \ll 1$. Some functionalities of thermal electric devices
require that the applied temperature bias $\Delta T$ violates the
$\Delta T/T \ll 1$ condition. In the following study, the
thermoelectric properties of QD junctions are investigated in the
nonlinear response regime.

\section{Nonlinear regime}

Scheibner and coworkers experimentally reported the thermal power of
the two-dimensional electron gas in QD under high magnetic fields in
the linear response regime.$^{27}$ Few theoretical works have
reported the thermal properties of QD junctions in the nonlinear
response regime.$^{28}$ Ref. 28 theoretically studied the thermal
power in the Kondo regime based on one-level Anderson model. Here,
we study the thermal electric effect of multiple QD junction in the
Coulomb blockade effect in the nonlinear regime. We show that in the
nonlinear regime, the thermal rectification behavior can become
quite pronounced. Records of thermal rectification date back to 1935
when Starr discovered that copper oxide/copper junctions can display
a thermal diode behavior.$^{29}$ Recently, thermal rectification
effects have been predicted to occur in one dimensional phonon
junction systems.$^{30-34}$

To study the direction-dependent heat current, we let
$T_L=T_0+\Delta T/2$ and $T_R=T_0-\Delta T/2$, where
$T_0=(T_L+T_R)/2$ is the equilibrium temperature of two side
electrodes and $\Delta T=T_L-T_R$ is the temperature difference.
Because the electrochemical potential difference, $e\Delta V$
yielded by the thermal gradient could be significant, it is
important to keep track the shift of the energy level of each dot
according to $\epsilon_{\ell}=E_{\ell}+\eta_{\ell}\Delta V/2$, where
$\eta_{\ell}$ is the ratio of the distance between dot $\ell$ and
the mid plane of the QD junction to the junction width. Here we set
$\eta_B=\eta_C=0$. A functional thermal rectifier requires a good
thermal conductor for $\Delta T > 0 $, but a poor thermal conductor
for $\Delta T < 0 $. Based on Eqs. (2) and (3), the asymmetrical
behavior of heat current with respect to $\Delta T$ requires not
only highly asymmetric coupling strengthes between the QDs and the
electrodes but also strong electron Coulomb interactions between
dots. To investigate the thermal rectification behavior, we have
numerically solved Eqs. (2) and (3) for multiple-QD junctions
involving two QDs and three QDs for various system parameters. We
first determine $\Delta V$ by solving Eq. (2) with  $J_e=0$ (the
open circuit condition) for a given $\Delta T$, $T_0$ and an initial
guess of the average one-particle and two-particle occupancy
numbers, $N_{\ell}$ and $c_{\ell}$ for each QD. Those numbers are
then updated according to Eqs. (5) and (6) until self-consistency is
established. Once $\Delta V$ is solved, we then use Eq. (3) to
compute the heat current.

Fig. 9 shows the heat currents, occupation numbers, and differential
thermal conductance (DTC) for the two-QD case, in which the energy
levels of dot A and dot B are $E_A=E_F-\Delta E/5$ and
$E_B=E_F+\alpha_B\Delta E$, where $\alpha_B$ is tuned between 0 and
1. We have adopted $\Delta E=200\Gamma$, which is used to describe
the energy level fluctuation of QDs. The heat currents are expressed
in units of $Q_0=\Gamma^2/(2h)$ through out this article. The
intradot and interdot Coulomb interactions used are
$U_{\ell}=30k_BT_0$ and $U_{AB}=15k_BT_0$. The tunneling rates are
$\Gamma_{AR}=0$, $\Gamma_{AL}=2\Gamma$, and
$\Gamma_{BR}=\Gamma_{BL}=\Gamma$. $k_BT_0$ is chosen to be
$25\Gamma$ throughout this article. Here, $\Gamma
=(\Gamma_{AL}+\Gamma_{AR})/2$ is the average tunneling rate in
energy units, whose typical values of interest are between 0.1 and
0.5 meV. The dashed curves are obtained by using a simplified
expression of Eq. (3) in which we set the average two particle
occupation in dots A and B to zero (resulting from the large
intradot Coulomb interactions) and taking the limit that $\Gamma \ll
k_BT_0$ so the Lorentzian function of resonant channels can be
replaced by a delta function. We have
\begin{eqnarray}
{Q}/\gamma_B &=& \pi (1-N_B)[(1-2N_A)(E_B-E_F)f_{LR}(E_B)
\nonumber \\ &+&2N_A(E_B+U_{AB}-E_F)f_{LR}(E_B+U_{AB})],
\end{eqnarray}
Here $N_{A(B)}$ is the average occupancy in dot A(B). Therefore,
it is expected that the curve corresponding to $E_B=E_F+4\Delta
E/5$ obtained with this delta function approximation is in good
agreement with the full solution, since $E_B$ is far away from the
Fermi energy level. For cases when $E_B$ is close to $E_F$, the
approximation is not as good, but it still gives qualitatively
correct behavior. Thus, it is convenient to use this simple
expression to illustrate the thermal rectification behavior. The
asymmetrical behavior of $N_A$ with respect to $\Delta T$ is
mainly resulted from the condition $\Gamma_{AR}=0$ and
$\Gamma_{AL}=2\Gamma$. The heat current is contributed from the
resonant channel with $\epsilon=E_B$, because the resonant channel
with $\epsilon=E_B+U_{AB}$ is too high in energy compared with
$E_F$. The sign of $Q$ is determined by $f_{LR}(E_B)$, which
indirectly depends on Coulomb interactions, tunneling rate ratio
and QD energy levels. The rectification behavior of $Q$ is
dominated by the factor $1-2N_A$, which explains why the energy
level of dot-A should be chosen below $E_F$ and the presence of
interdot Coulomb interactions is crucial. The negative sign of Q
in the regime of $\Delta T < 0$ indicates that the heat current is
from the right electrode to the left electrode. We define the
rectification efficiency as $\eta_{Q}=(Q(+\Delta T)-|Q(-\Delta
T)|)/Q(+\Delta T)$. We obtain $\eta_{Q}(\Delta T=30\Gamma)=0.86$
for $E_B=E_F+2\Delta E/5$ and $\eta_Q(\Delta T=30\Gamma)=0.88$ for
$E_B=E_F+4\Delta E/5$.  Fig. 9(c) shows DTC in units of $Q_0
k_B/\Gamma$. It is found that the rectification behavior is not
very sensitive to the variation of $E_B$. DTC is roughly linearly
proportional to $\Delta T$ in the range $-20\Gamma < k_B \Delta T
< 20 \Gamma$. In addition, we also find a small negative
differential thermal conductance (NDTC) for $E_B=E_F+4\Delta E/5$.
Similar behavior was reported in the phonon junction
system.$^{35}$

Note that the mechanism of thermal rectification is similar to the
charge current rectification. However, the heat current is yielded
by the temperature bias and the electrochemical potential. In
particular, the electrochemical potential is a highly nonlinear
function of the temperature bias, which has never been reported for
quantum dot junctions. Consequently, it is not straightforward to
reveal the behavior of heat current with respect to the
temperature bias. The manifested difference between the heat
current and the charge current is that the origin of NDTC is
different from that of negative differential conductance (NDC).
The NDC of charge current requires the upper energy levels with
the shell-filling condition, which was discussed in
Refs. 18 and 19. For NDTC, it only appears in the lower level
with shell-filling condition. Fig. 10 shows the rectification
efficiency as a function $\Delta T$ for two different values of
$E_B$. The rectification efficiency vanishes when $k_B\Delta
T/\Gamma \ll 1$. This implies that it is difficult to judge the
rectification effect in the linear response regime of $\Delta
T/T_0 \ll 1$. Although the two-dot case can reach a high
rectification efficiency, the heat current should be enhanced from
the application point of view.

Fig. 11 shows the heat current, differential thermal conductance
and thermal power as functions of temperature difference $\Delta
T$ for a three-QD case for various values of $\Gamma_{AR}$, while
keeping $\Gamma_{B(C),R}=\Gamma_{B(C),L}=\Gamma$. Here, we adopt
$\eta_A=|\Gamma_{AL}-\Gamma_{AR}|/(2\Gamma)$ instead of fixing
$\eta_A$ at 0.3 to reflect the correlation of dot position with
the asymmetric tunneling rates. We assume that the three QDs are
roughly aligned with dot A in the middle. The energy levels of
dots A, B and C are chosen to be $E_A=E_F-\Delta E/5$,
$E_B=E_F+2\Delta E/5$ and $E_C=E_F+3\Delta E/5$.
$U_{AC}=U_{BA}=15k_BT_0$, $U_{BC}=8k_BT_0$, $U_C=30 k_BT_0$, and
all other parameters are kept the same as in the two-dot case. The
thermal rectification effect is most pronounced when
$\Gamma_{AR}=0.$ as seen in Fig. 12(a). (Note that the heat
current is not very sensitive to $U_{BC}$). In this case, we
obtain a small heat current $Q=0.068 Q_0$ at $\Delta T=-30\Gamma$,
but a large heat current $Q=0.33 Q_0$ at $\Delta T =30 \Gamma$ and
the rectification efficiency $\eta_{Q}$ is 0.79. However, the heat
current for $\Gamma_{AR}=0$ is small. For $\Gamma_{AR}=0.1\Gamma$,
we obtain $Q=1.69 Q_0$ at $\Delta T=-30\Gamma$,$Q=5.69 Q_0$ at
$\Delta T =30 \Gamma$, and $\eta_{Q}= 0.69$. We see that the heat
current is  suppressed for $\Delta T < 0$ with decreasing
$\Gamma_{AR}$. This implies that it is important to blockade the
heat current through dot A to observe the rectification effect.
Very clear NDTC is observed in Fig. 11(b) for the
$\Gamma_{AR}=0.1\Gamma$ case, while DTC is symmetric with respect
to $\Delta T$ for the $\Gamma_{AR}=\Gamma_{AL}$ case.

From the experimental point of view, it is easier to measure the
thermal power than the direction-dependent heat current. The thermal
power as a function of $\Delta T$ is shown in Fig. 11(c). All curves
except the dash-dotted line (which is for the symmetrical tunneling
case) show highly asymmetrical behavior with respect to $\Delta T$,
yet it is not easy at all to judge the efficiency of the
rectification effect from $S$ for small $|\Delta T|$ ($k_B|\Delta
T|/\Gamma < 10$). Thus, it is not sufficient to determine whether a
single QD can act as an efficient thermal rectifier based on results
obtained in the linear response regime of $\Delta T/T_0 \ll
1$.$^{27}$ According to the thermal power values, the
electrochemical potential $e\Delta V$ can be very large.
Consequently, the shift of QD energy levels caused by $\Delta V$ is
quite important. To illustrate the importance of this effect, we
plot in Fig. 12 the heat current for various values of $E_C$ for the
case with $\Gamma_{AR}=0$, $U_{BC}=10k_BT_0$ and $\eta_A=0.3$. Other
parameters are kept the same as those for Fig. 11. The solid
(dashed) curves are obtained by including (excluding) the energy
shift $\eta_A \Delta V/2$. It is seen that the shift of QD energy
levels due to $\Delta V$ can lead to significant change in the heat
current. It is found that NDTC is accompanied with low heat current
for the case of $E_C=E_F+\Delta E/5$ [see Fig. 12(b)]. Even though
the heat current exhibits rectification effect for $E_C=E_F+\Delta
E/5$ and $E_C=E_F+3\Delta E/5$, the thermal power has a very
different behavior. From Figs. 11(c) and 12(c), we see that the heat
current is a highly nonlinear function of electrochemical potential,
$\Delta V$. Consequently, the rectification effect is not
straightforwardly related to the thermal power in this system.

Because the position distribution fluctuation is common for QDs,
we investigate the interdot Coulomb interactions on the
rectification effect. Fig. 13 shows the heat current,
electrochemical potential and occupation number as functions of
$\Delta T$ for various values of $U_{AC}$ with $E_C=E_F+\Delta E/5$.
Other parameters are the same as in Fig. 12. When $U_{AC}=0$, the
rectification efficiency is suppressed seriously. The residue
rectification mainly arises from the correlation between dot A and
dot B. Such results indicate that it is crucial to control the QD
position in the implementation of QD thermal rectifiers. We find
that the electrochemical potential is not significantly changed
when $U_{AC}$ decreases, whereas the heat current has a
considerable variation. Fig. 13(c) shows the occupation numbers of
dots A and C. $N_B$ are ignored due to their energy levels being far away
from the Fermi energy level. It is expected that $N_A$ is not
sensitive to the decrease of $U_{AC}$. $N_C$ increases so much
when $U_{AC}$ decreases since the main resonant channels of dot C
are dominated by $E_C$ rather than the combination of $E_C$ and
$E_C+U_{AC}$. Fig. 13(c) reveals that the serious suppression of
rectification efficiency of dot-dashed line shown in Fig. 13(a) is
mainly attributed to the heat current through dot C. We once again
investigate the rectification efficiency for three-dot case. Fig.
14 shows the rectification efficiency as function of $\Delta T$.
All other parameters  are the same as those of Fig. 12. The
rectification efficiency increases with  increasing
temperature bias. However, $\eta_Q$ is not sensitive to
the energy level of dot C.

Comparing the heat current of the three-dot case (shown in Figs.
11 and 12) to the two-dot case (shown in Fig. 9), we find that the
rectification efficiency is about the same for both cases (shown
in Figs. 10 and 14), while the magnitude of the heat current can
be significantly enhanced in the three-dot case. For practical
applications, we need to estimate the magnitude of the heat
current density and DTC of the IQV junction device in order to see
if the effect is significant. We envision a thermal rectification
device made of an array of multiple QDs (e.g. three-QD cells) with
a 2D density $N_{2d} = 10^{11} cm^{-2}$. For this device, the heat
current density versus $\Delta T$ is given by Figs. 11 and 12 with
the units $Q_0$ replaced by $N_{2d}Q_0$, which is approximately
$965 {} W/m^2$ if we assume $\Gamma=0.5 meV$. Similarly, the units
for DTC becomes $N_{2d}k_B Q_0 /\Gamma$, which is approximately
$34 {} W/m^2{}^0$K. Since the phonon contribution can be blocked
by the vacuum layer in our design, this device could have
practical applications near $140^0$K with ($k_BT_0 \approx 12.5
meV$). If we choose a higher tunneling rate $\Gamma > 1 meV$ and
Coulomb energy $>300 meV$ (possible for QDs with diameter less
than 1 nm), then it is possible to achieve room-temperature
operation. It is worth pointing out that if the vacuum layer is replaced by a
typical phonon glass, such as SiO$_2$, which has a thermal conductivity of $\kappa_{ph}=  1.5 W/m {}^0$K [36] at room temperature,
the heat current carried by phonons across a 10 nm junction with a temperature bias of 1${}^0$K would be around $1.5 \times 10^8 W/m^2$. This would completely dominate over the thermal electric effect considered here (by six orders of magnitude). Therefore, unless a vacuum layer is inserted, the term $\kappa_{ph}$ will play a dominant role.

\section{Summary and conclusions}


We have theoretically investigated the effect of intradot and interdot Coulomb interactions on the
figure of merit (ZT) and thermal power (S) of multiple QD junction
system in the sequential tunneling process. The ZT values at high
temperatures are significantly suppressed by the intradot as well
as interdot Coulomb interactions. The optimization of ZT depends
not only on temperature but also on the detuning energy ($\Delta=E_g-E_F$).
It is worth noting that inelastic scattering effect
arising from QD size fluctuations, defects and electron-phonon
interactions will lead to considerable reduction to the ZT values.
Electrical conductance and thermal power exhibit
Coulomb oscillatory behavior and the sawtooth-like behavior with
respect to the gate voltage. The largest value for ZT is obtained
midway between good and poor conductors. Apart from the
results of linear response, the heat rectification effect can be
observed for multiple QD junctions in the nonlinear response
regime. In contrast to the heat rectification of phonon junction
system, the heat current is carried by electrons in the multiple QD junction system and
large electrochemical potentials can be established by the temperature gradient to generate electrical power.

{\bf Acknowledgments}\\
This work was supported in part by the National Science Council of the
Republic of China under Contract Nos. NSC 97-2112-M-008-017-MY2
and NSC 98-2112-M-001-022-MY3 and by Academia Sinica.


\mbox{}\\
${}^{\dagger}$ E-mail address: mtkuo@ee.ncu.edu.tw\\
$^*$ E-mail address: yiachang@gate.sinica.edu.tw

\mbox{}

\newpage

{\bf Figure Captions}

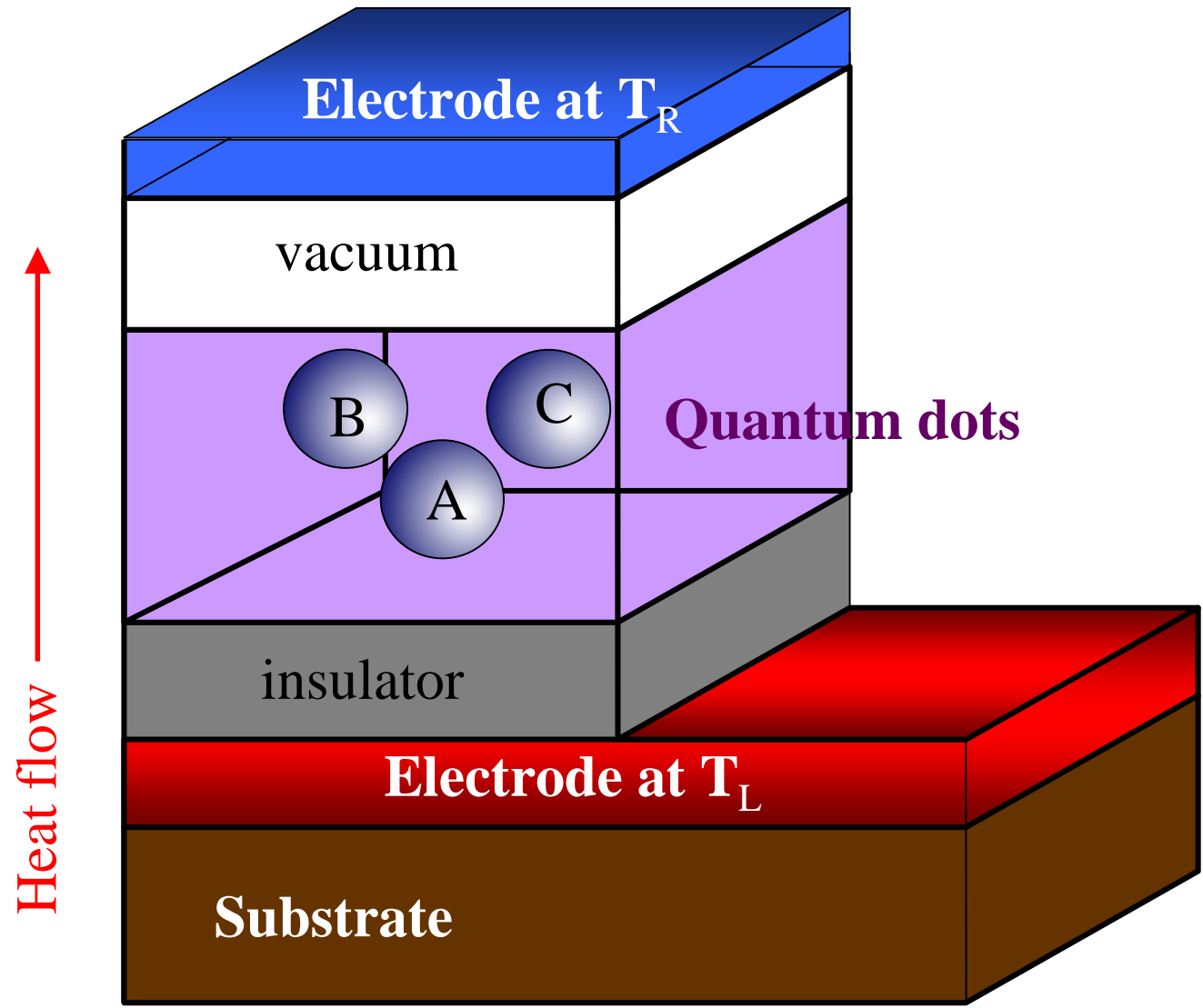
Fig. 1. Schematic diagram of the isulator/quantum dots/vacuum (IQV)
tunnel junction device.

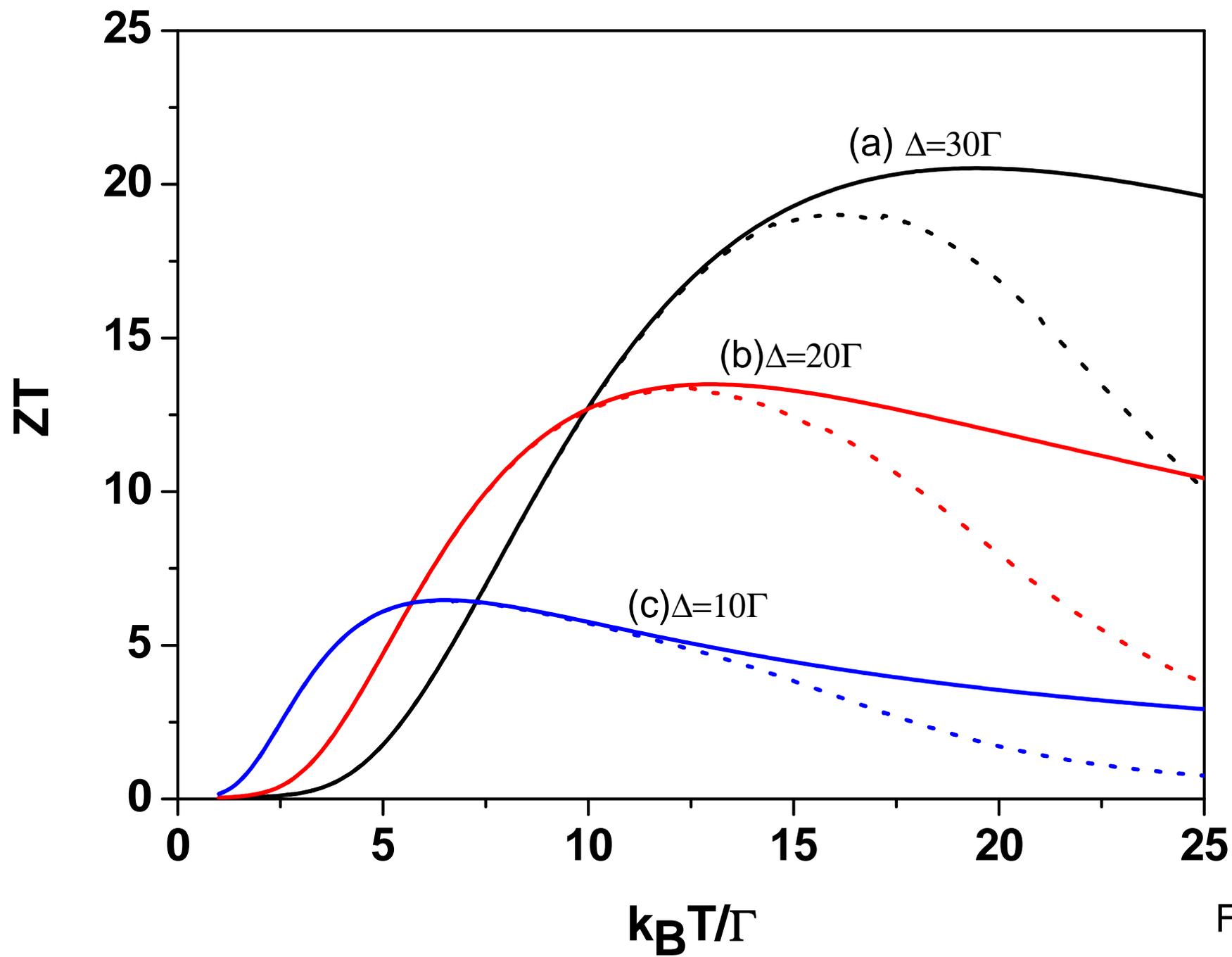
Fig. 2. Figure of merit ZT as a function of temperature for
various values of $\Delta $in the absence of interdot Coulomb
interactions.  Solid lines and dotted lines correspond to $U=0$
and $U=125\Gamma$, respectively.

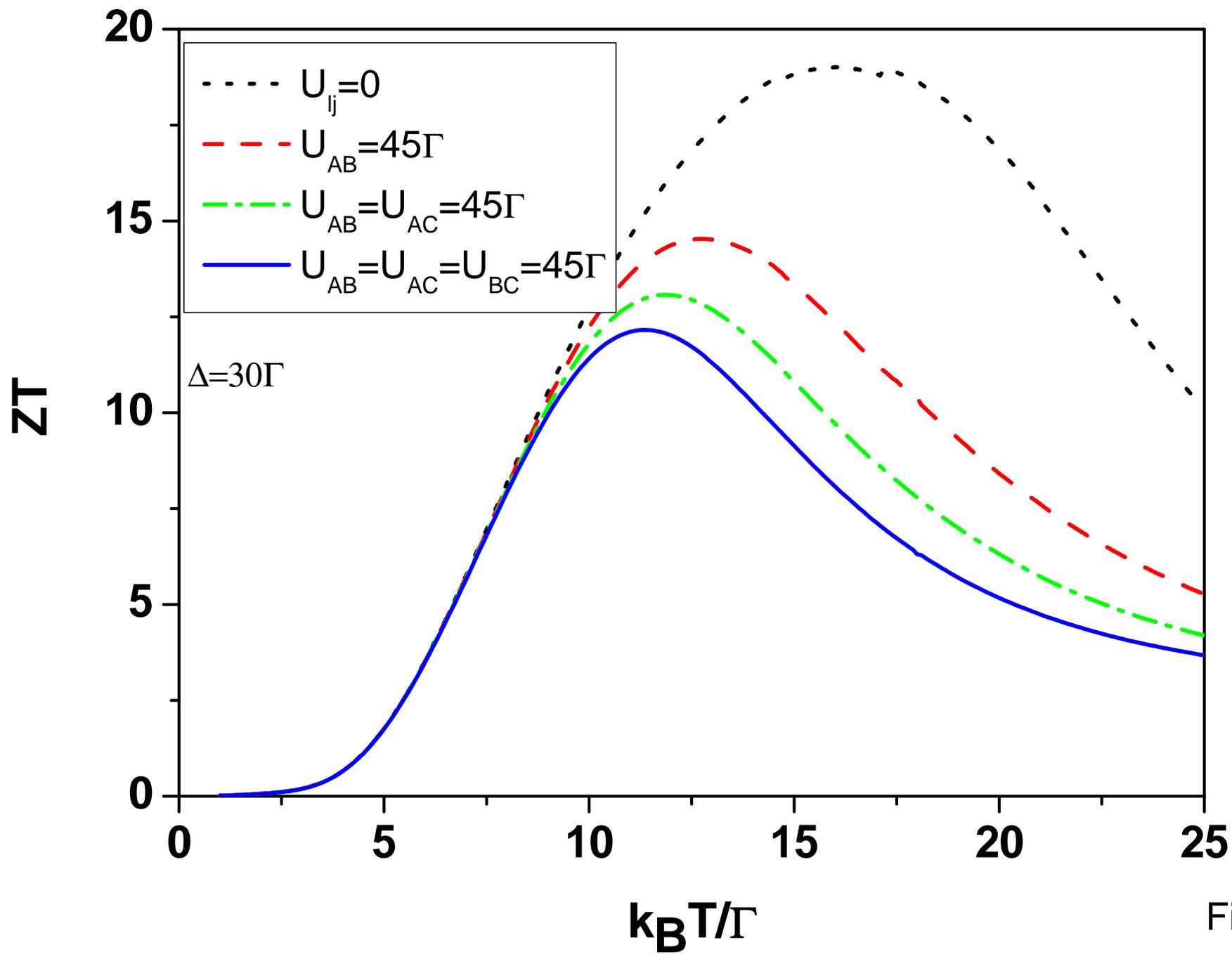
Fig. 3. Figure of merit ZT as a function of temperature for
different quantum dot configurations.

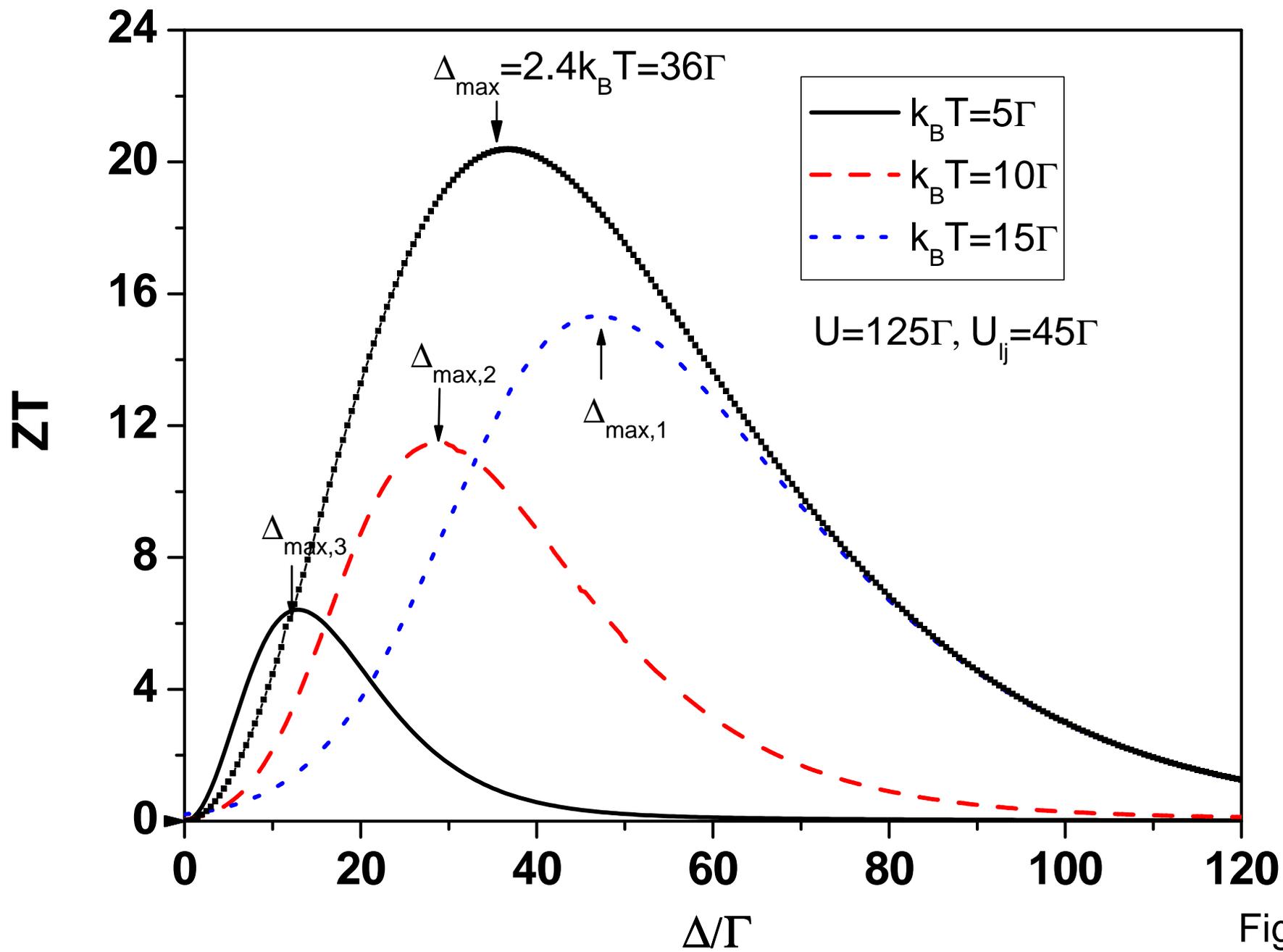
Fig. 4. Figure of merit ZT as a function of $\Delta$ for different
temperatures.

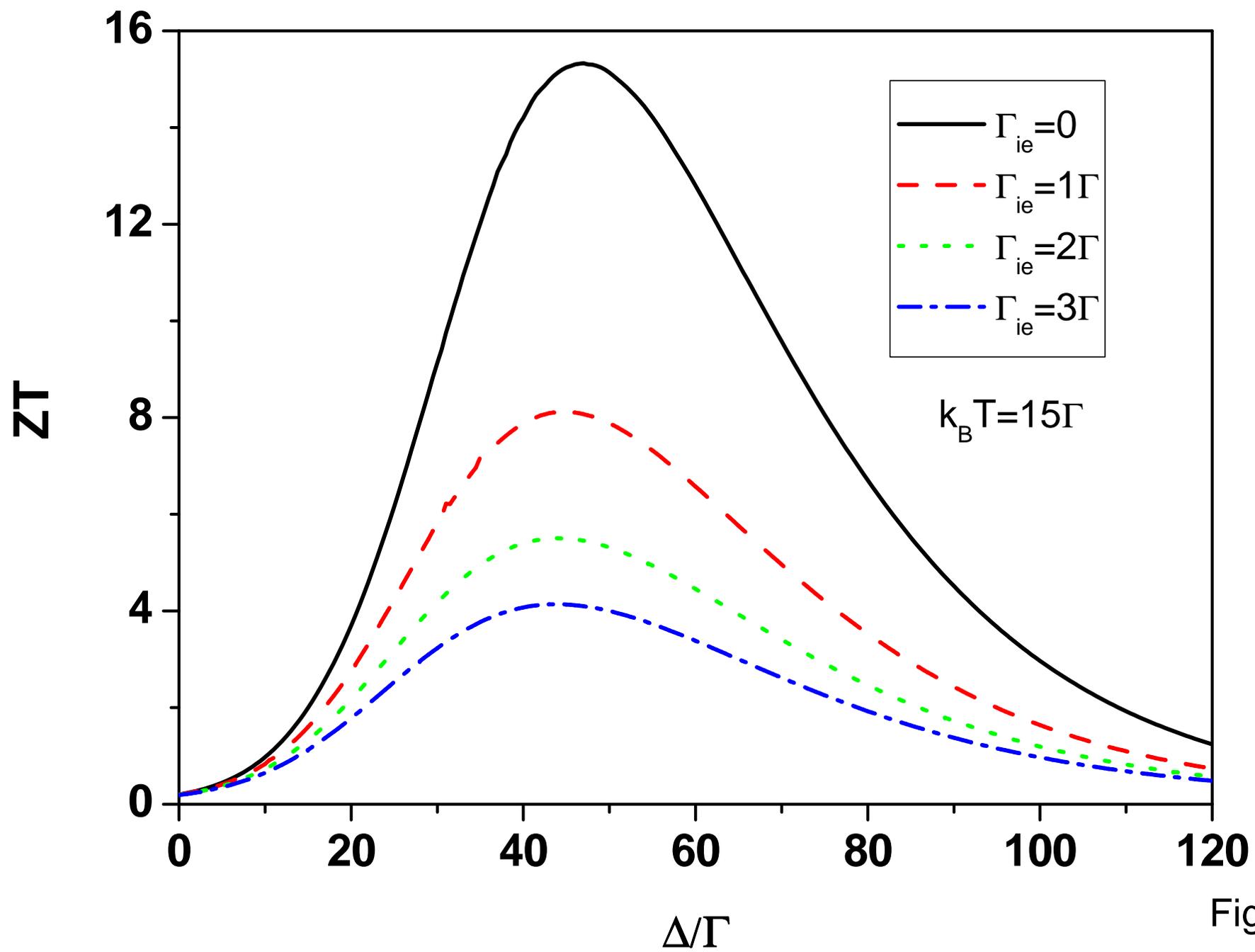
Fig. 5. Figure of merit ZT as a function of $\Delta$ for different
inelastic scattering strengthes at $k_BT=15\Gamma$.

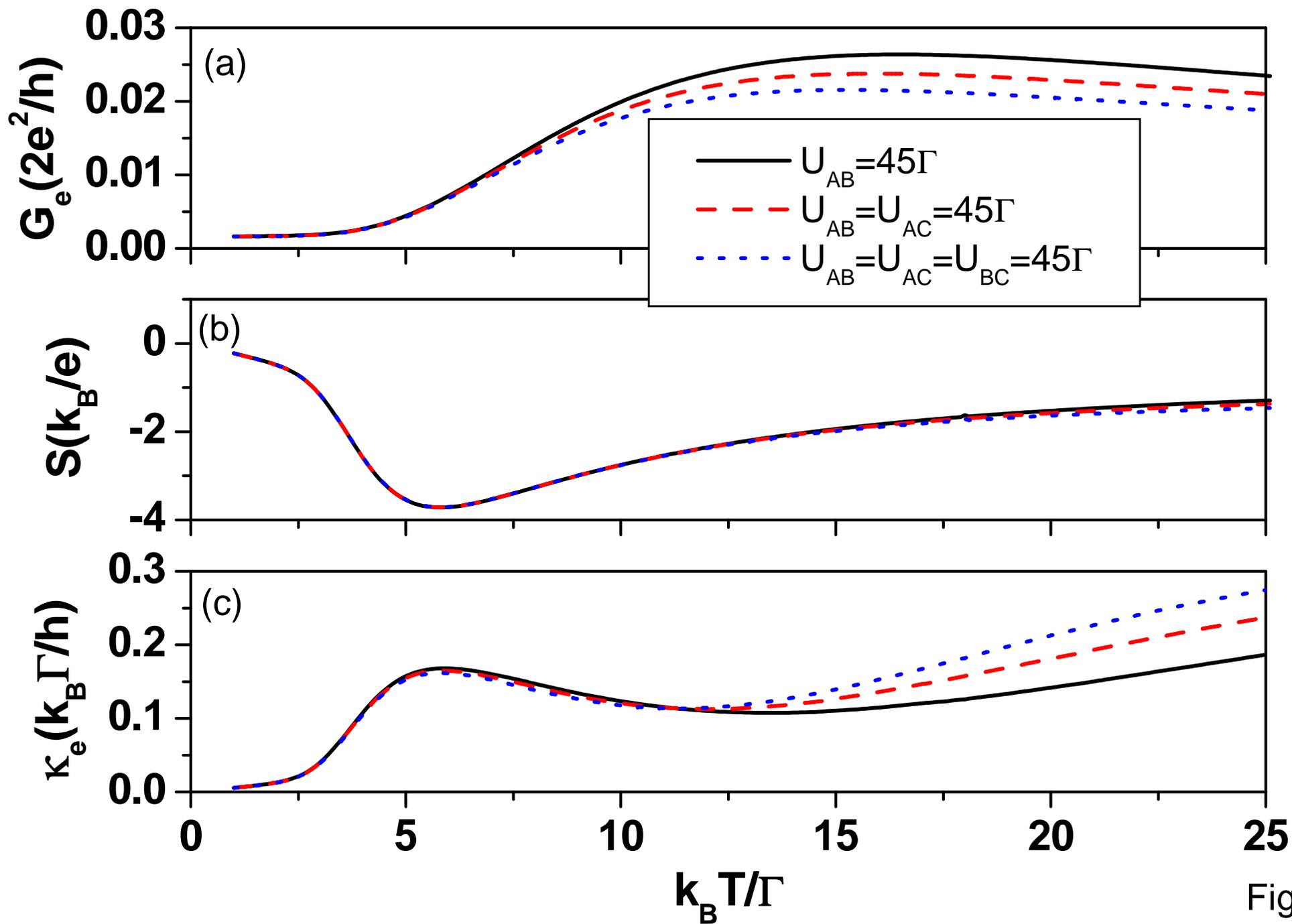
Fig. 6. Electrical conductance $G_e$, thermal power S and electron
thermal conductance $\kappa_e$ as a function of temperature for
different quantum dot configurations.

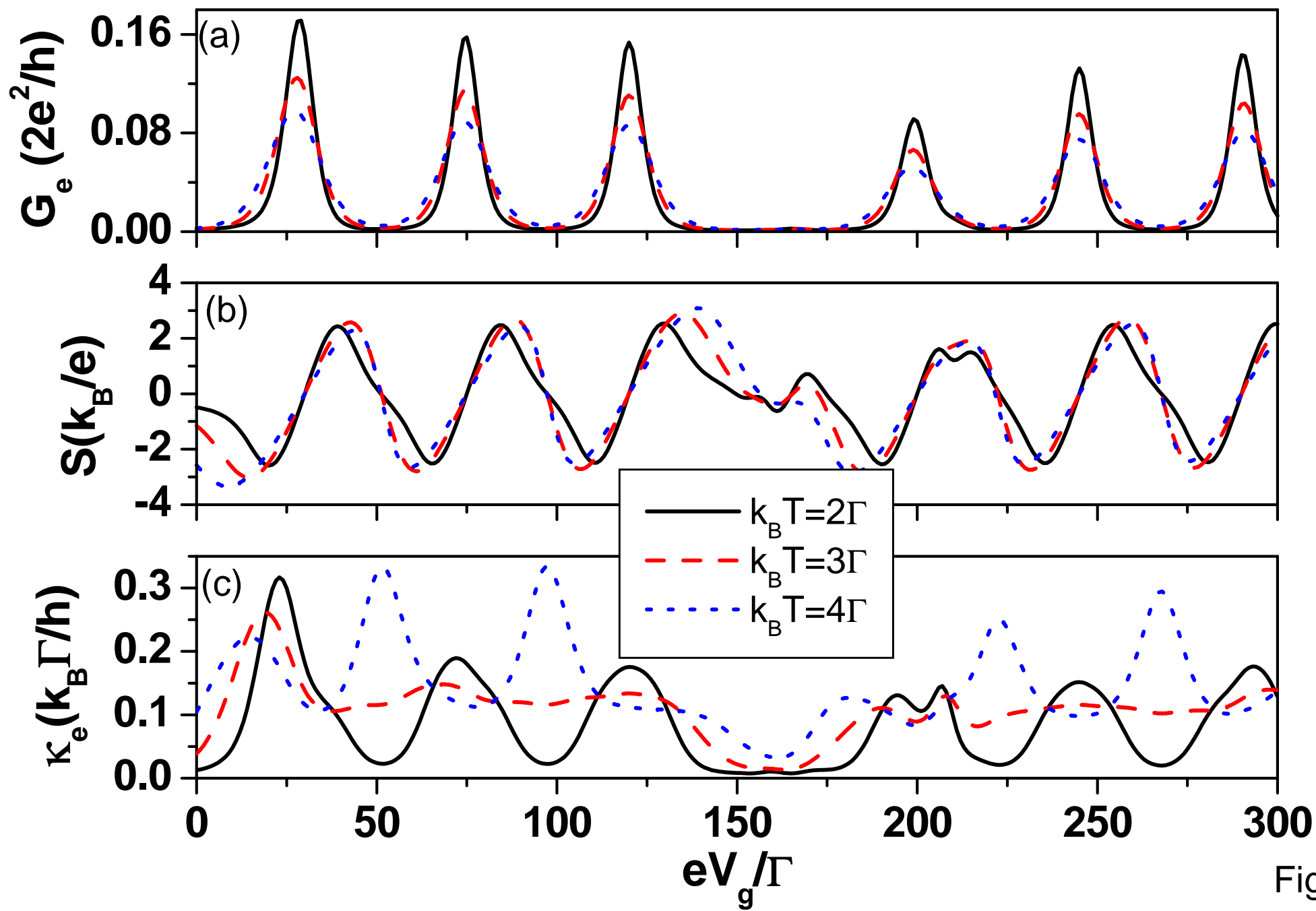
Fig. 7. Electrical conductance $G_e$, thermal power S and electron
thermal conductance $\kappa_e$ as a function of applied gate
voltage for different temperatures at $\Delta=30\Gamma$,
$U=125\Gamma$ and $U_{lj}=45\Gamma$.

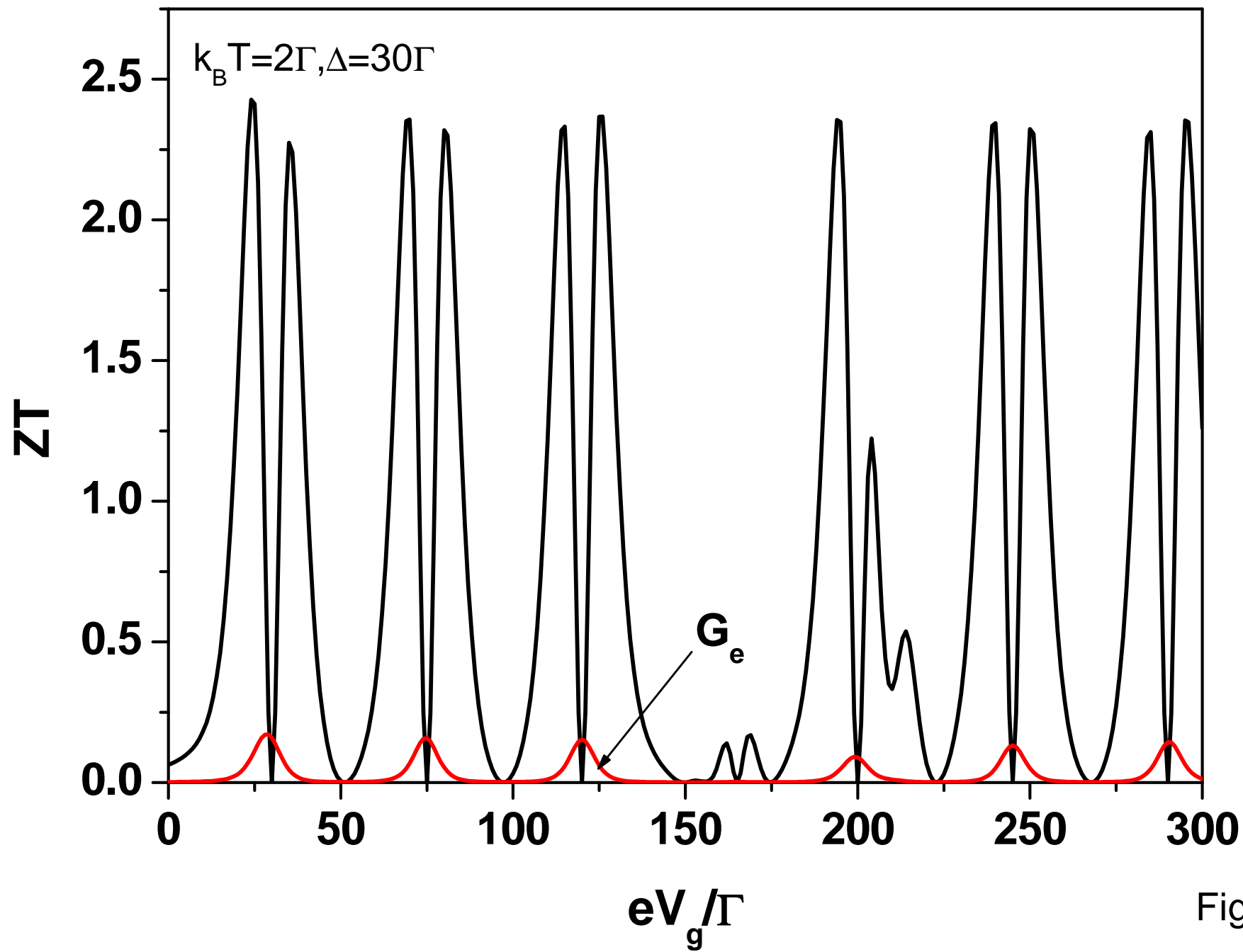
Fig. 8. Figure of merit as a function of applied gate voltage at
$k_BT=2\Gamma$, $\Delta=30\Gamma$, $U=125\Gamma$ and
$U_{lj}=45\Gamma$.

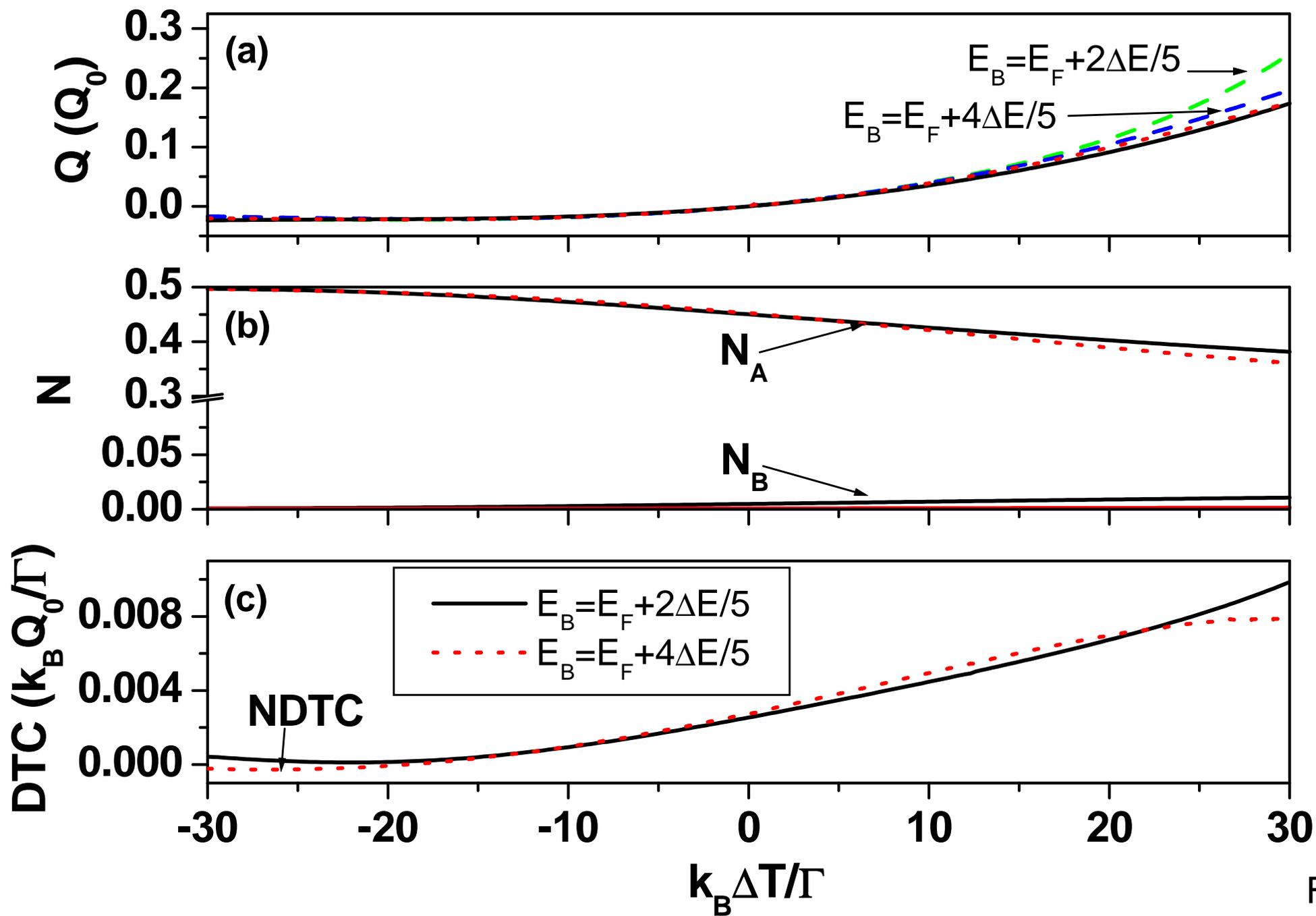
Fig. 9. (a) Heat current (b) average occupation number, and (c)
differential thermal conductance as a function of $\Delta T$ for
various values of $E_B$ for a two-QD junction. $\Gamma_{AR}=0$,
$\eta_A=0.3$ and $\Delta E= 200 \Gamma$.

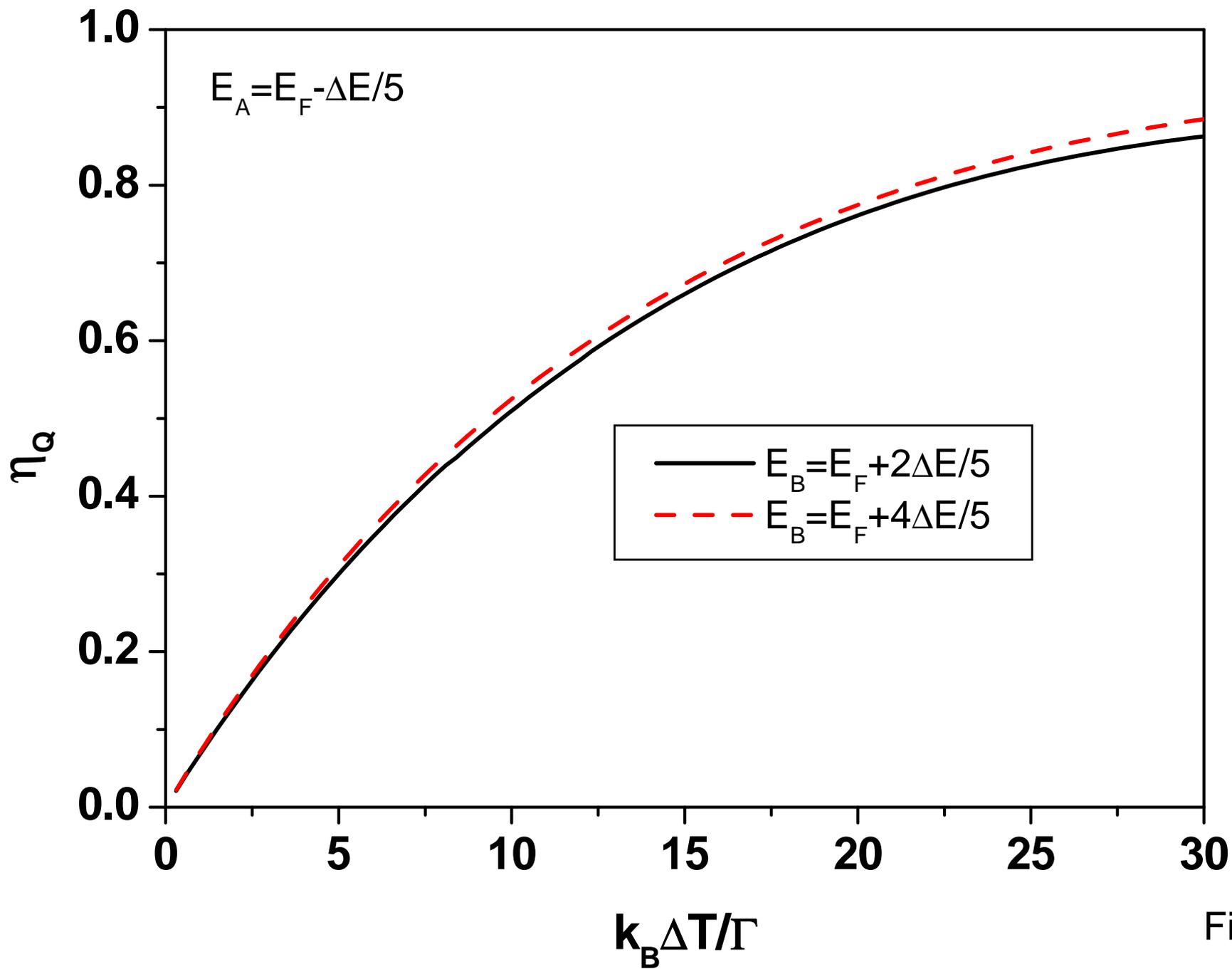
Fig. 10. Rectification efficiency as a function of $\Delta T$ for
two different values of $E_B$. Other parameters are the same as
those of Fig. 9.

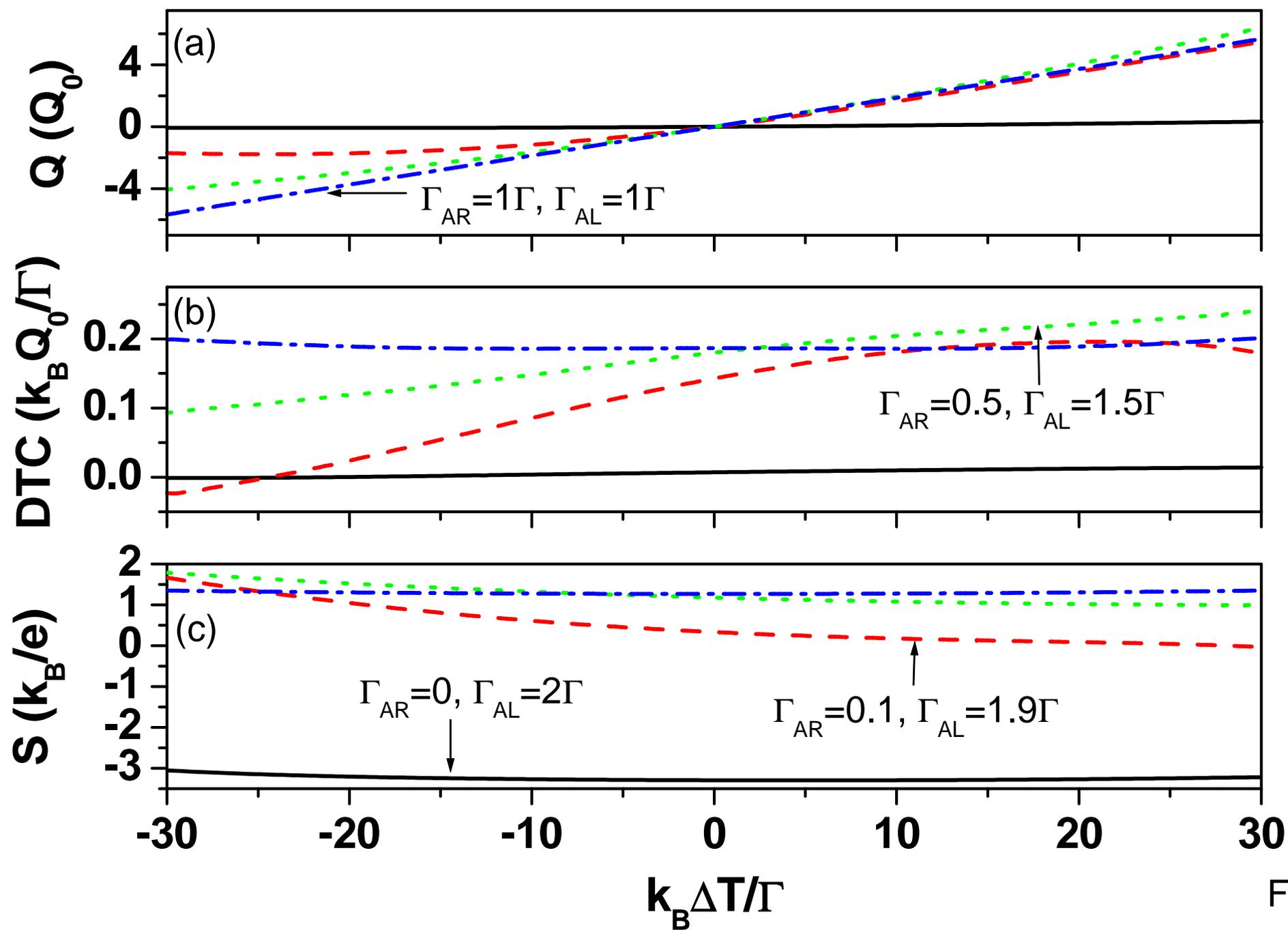
Fig. 11. (a) Heat current, (b) differential thermal conductance
and (c) thermal power as a function of $\Delta T$ for various
values of $\Gamma_{AR}$ for a three-QD junction.

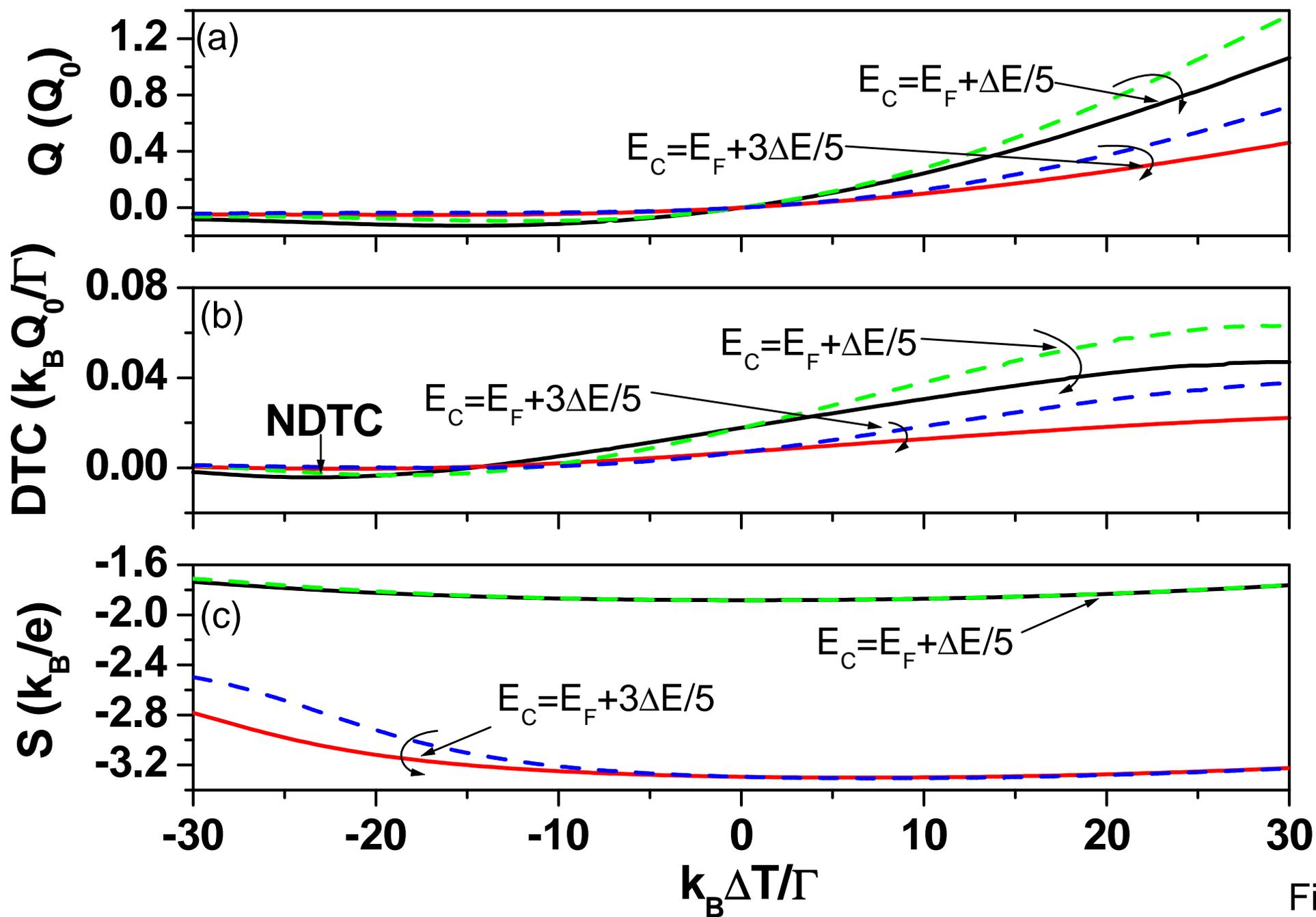
Fig. 12. (a) Heat current, (b) differential thermal conductance
and (c) thermal power as functions of $\Delta T$ for various
values of $E_C$ for a three-QD junction with $\Gamma_{AR}=0$ and
$\eta_A=0.3$.

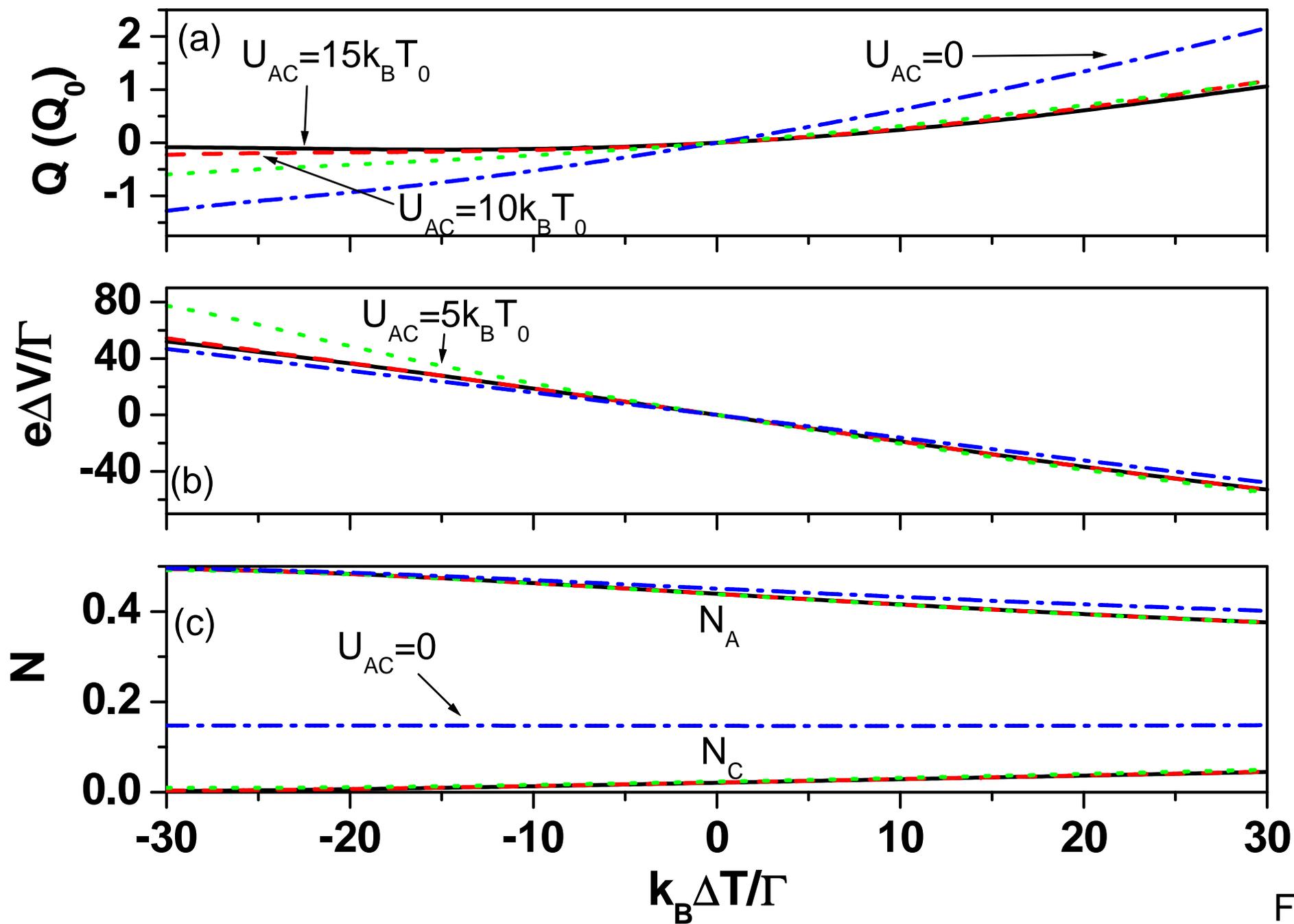
 Fig. 13. (a) Heat current, (b) electrochemical potential and (c)
 occupation number as a function of $\Delta T$ for various values
 of $U_{AC}$ for  a three-QD junction with $E_C=E_F+\Delta E/5$.
 All other parameters are same as in Fig. 12.

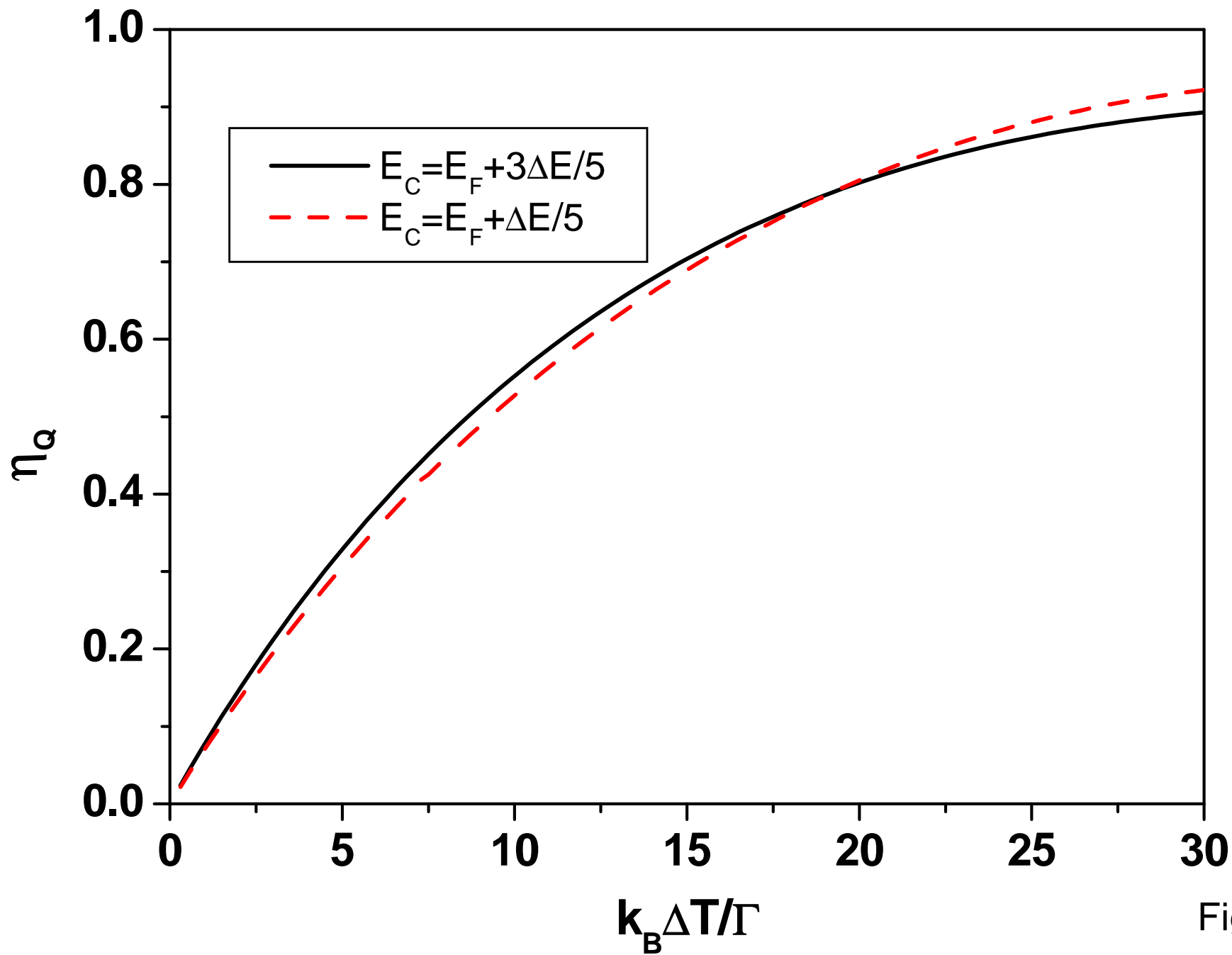
 Fig. 14. Rectification efficiency as a function of $\Delta T$ for
two different values of $E_C$. Other parameters are the same as
those of Fig. 12.


\begin{thebibliography}{50}

\bibitem[1]{Min} A. J. Minnich, M. S. Dresselhaus, Z. F. Ren and
G. Chen, Energy Environ Sci, \textbf{2}, 466 (2009).

\bibitem[2]{Mah} G. Mahan, B. Sales and J. Sharp, Physics Today,
\textbf{50},  42 (1997).

\bibitem[3]{Ven} R. Venkatasubramanian, E. Siivola,T. Colpitts,B. O'Quinn, Nature  \textbf{413},
597  (2001).

\bibitem[4]{Bou} A. I. Boukai, Y. Bunimovich, J. Tahir-Kheli, J. K.
Yu, W. A. Goddard III and J. R. Heath, Nature, \textbf{451}, 168
(2008).

\bibitem[5]{Har} T. C. Harman, P. J. Taylor, M. P. Walsh, B. E.
LaForge, Science \textbf{297},  2229 (2002).


\bibitem[6]{Hsu} K. F. Hsu,S. Loo,F. Guo,W. Chen,J. S. Dyck,C. Uher, T. Hogan,
E. K. Polychroniadis,M. G. Kanatzidis, Science \textbf{303}, 818
(2004).

\bibitem[7]{Maj} A. Majumdar, Science \textbf{303},  777 (2004).

\bibitem[8]{Che1} G. Chen, M. S. Dresselhaus, G. Dresselhaus, J. P.
Fleurial and T. Caillat, International Materials Reviews,
\textbf{48},  45 (2003).

\bibitem[9]{Lin} Y. M. Lin and M. S. Dresselhaus, Phys. Rev. B \textbf{68},
 075304 (2003).


\bibitem[10]{Mur} P. Murphy, S. Mukerjee and J. Moore, Phys. Rev.
B \textbf{78}, 161406 (2008).

\bibitem[11]{Win} N. S. Wingreen, K. W. Jacobsen and J. W. Wikins,
Phys. Rev. B \textbf{40}, 11834 (1989).

\bibitem[12]{Jau} A. P. Jauho, N. S. Wingreen and Y. Meir,
Phys. Rev. B \textbf{50}, 5528 (1994).

\bibitem[13]{Kuo1} D. M. T. Kuo and Y. C. Chang, Phys. Rev. B \textbf{66},
085311 (2002).

\bibitem[14]{Lun} U. Lundin and R. H. McKenzie, Phys. Rev. B
\textbf{66}, 075303 (2002).

\bibitem[15]{Fle} K. Flensberg, Phys. Rev. B \textbf{68}, 205323 (2003).


\bibitem[16]{Che} Z. Z. Chen, R. Lu and B. F. Zhu, Phys. Rev. B
\textbf{71}, 165324 (2005).

\bibitem[17]{Gal} M. Galperin, A. Nitzan and M. A. Ratner, Phy. Rev. B \textbf{75},
155312 (2007).

\bibitem[18]{Kuo} D. M. T. Kuo and Y. C. Chang, Phys. Rev. Lett.
\textbf{99}, 086803  (2007).

\bibitem[19]{Cha} Y. C. Chang and D. M. T. Kuo, Phys. Rev. B
\textbf{77}, 245412 (2008).

\bibitem[20]{Wu} L. A. Wu and D. Segal, Phys. Rev. Lett. {\bf
102},  095503 (2009).

\bibitem[21]{Sai} O. P. Saira, M. Meschke, F. Giazotto, A. M.
Savin, M. Mottonen and J. Pekola, Phys. Rev. Lett. \textbf{99},
027203 (2007).

\bibitem[22]{Hag} H. Haug and A. P. Jauho, \emph{Quantum Kinetics in Transport and
Optics of Semiconductors }(Springer, Heidelberg, 1996).

\bibitem[23]{Dat}  S. Datta, Electronic Transport in Mesoscopic
Systems (Cambridge University Press, Cambridge U. K. (1995)).

\bibitem[24]{Sch1} R. Scheibner, E. G. Novik, T. Borzenko, M.
Konig, D. Reuter, A. D. Wieck, H. Buhmann, and L. W. Molenkamp,
Phys. Rev. B \textbf{75}, 041301 (2007).

\bibitem[25]{Zia} X. Zianni, Phys. Rev. B \textbf{78}, 165327 (2008).

\bibitem[26]{Bee} C. W. J. Beenakker: Phys. Rev. B \textbf{44},
1646 (1991).

\bibitem[27]{Sch} R. Scheibner, M. Konig, D. Reuter, A. D. Wieck, C.
Gould, H. Buhmann and L. W. Molenkamp,  New. J. Phys. \textbf{10},
083016 (2008).

\bibitem[28]{Kra} M. Krawiec and K. I. Wysokinski, Phys. Rev. B \textbf{75},
155330 (2007).

\bibitem[29]{Sta} C. Starr, J. Appl. Phys. \textbf{7}, 15 (1936).

\bibitem[30]{Ter} M. Terraneo, M. Peyrard, G. Casati, Phys. Rev.
Lett. \textbf{88}, 094302 (2002).

\bibitem[31]{Li} Baowen Li, L. Wang and G. Casati, Phys. Rev. Lett.
\textbf{93}, 184301 (2004).

\bibitem[32]{Hu} B. Hu, L. Yang and Y. Zhang, Phys. Rev. Lett. \textbf{97},
124302 (2006).

\bibitem[33]{Cas} G. Casati, C. Mejia-Monasterio and T. Prosen,
Phys. Rev. Lett. \textbf{98}, 104302 (2007).

\bibitem[34]{Zen} N. Zeng and J. S. Wang, Phys. Rev. B \textbf{78}, 024305
(2008).


\bibitem[35]{Seg} D. Segal, Phys. Rev. B \textbf{73}, 205415 (2006).

\bibitem[36]{Ca} D. G. Cahill and R. O. Pohl, Phys. Rev. B \textbf{35}, 4067 (1987).









\end{thebibliography}
\end{document}